\documentclass[sn-mathphys,Numbered]{sn-jnl}

\usepackage{graphicx}%
\usepackage{multirow}%
\usepackage{amsmath,amssymb,amsfonts}%
\usepackage{amsthm}%
\usepackage{mathrsfs}%
\usepackage[title]{appendix}%
\usepackage{xcolor}%
\usepackage{textcomp}%
\usepackage{manyfoot}%
\usepackage{booktabs}%
\usepackage{algorithm}%
\usepackage{algorithmicx}%
\usepackage{algpseudocode}%
\usepackage{listings}%
\usepackage{placeins}

\begin{document}

\title[Optimizing High Throughput Inference on Graph Neural Networks at Shared Computing Facilities with the NVIDIA Triton Inference Server]{Optimizing High Throughput Inference on Graph Neural Networks at Shared Computing Facilities with the NVIDIA Triton Inference Server}


\author*[1]{\fnm{Claire} \sur{Savard}}\email{Claire.Savard@colorado.edu}

\author[1]{\fnm{Nicholas} \sur{Manganelli}}\email{Nicholas.James.Manganelli@cern.ch}

\author[2]{\fnm{Burt} \sur{Holzman}}\email{burt@fnal.gov}

\author[2]{\fnm{Lindsey} \sur{Gray}}\email{Lindsey.Gray@cern.ch}

\author[1,3]{\fnm{Alexx} \sur{Perloff}}\email{perloff1@llnl.gov}

\author[2]{\fnm{Kevin} \sur{Pedro}}\email{pedrok@fnal.gov}

\author[1]{\fnm{Kevin} \sur{Stenson}}\email{Kevin.Stenson@colorado.edu}

\author[1]{\fnm{Keith} \sur{Ulmer}}\email{Keith.Ulmer@colorado.edu}

\affil[1]{\orgdiv{Department of Physics}, \orgname{University of Colorado, Boulder}, \orgaddress{\street{390 UCB}, \city{Boulder}, \postcode{80309}, \state{CO}, \country{USA}}}

\affil[2]{\orgname{Fermi National Accelerator Laboratory}, \orgaddress{\street{Wilson Street and Kirk Road}, \city{Batavia}, \postcode{60510}, \state{IL}, \country{USA}}}

\affil[3]{\orgname{Lawrence Livermore National Laboratory}, \orgaddress{\street{7000 East Avenue}, \city{Livermore}, \postcode{94550}, \state{CA}, \country{USA}}}


\abstract{
With machine learning applications now spanning a variety of computational tasks, multi-user shared computing facilities are devoting a rapidly increasing proportion of their resources to such algorithms. Graph neural networks (GNNs), for example, have provided astounding improvements in extracting complex signatures from data and are now widely used in a variety of applications, such as particle jet classification in high energy physics (HEP). However, GNNs also come with an enormous computational penalty that requires the use of GPUs to maintain reasonable throughput. At shared computing facilities, such as those used by physicists at Fermi National Accelerator Laboratory (Fermilab), methodical resource allocation and high throughput at the many-user scale are key to ensuring that resources are being used as efficiently as possible. These facilities, however, primarily provide CPU-only nodes, which proves detrimental to time-to-insight and computational throughput for workflows that include machine learning inference. In this work, we describe how a shared computing facility can use the NVIDIA Triton Inference Server to optimize its resource allocation and computing structure, recovering high throughput while scaling out to multiple users by massively parallelizing their machine learning inference. To demonstrate the effectiveness of this system in a realistic multi-user environment, we use the Fermilab Elastic Analysis Facility augmented with the Triton Inference Server to provide scalable and high throughput access to a HEP-specific GNN and report on the outcome.
}




\maketitle

\section{Introduction}

Machine learning (ML) is a continually growing field, gaining traction across disciplines as new applications are found and tested. In high energy physics (HEP), for example, ML frequently outperforms traditional algorithms, leading to adoption for a wide variety of tasks, now encompassing the reconstruction and classification of physics objects and events recorded by particle detectors such as those at the Large Hadron Collider (LHC)~\cite{ml_in_hep,dl_lhc}. The most powerful ML techniques, such as graph neural networks (GNNs), are more complex and correspondingly require more computing power and time~\cite{cpu_vs_gpu,bench_forDL}.

Computing power can be expensive and is not readily available to everyone. Therefore, many turn towards shared computing facilities that give users access to otherwise unaffordable computational resources~\cite{cluster_comp}. In general, these facilities provide a variety of different platforms and processors to users, such as CPUs and GPUs, but tend to be optimized for conventional tasks requiring minimal computational power per user. 
Facilities like the LHC Physics Center~\cite{lpc} at Fermi National Accelerator Laboratory (Fermilab), which serves Large Hadron Collider (LHC) physicists from the CMS experiment, provide resources to hundreds of HEP researchers per year, but now struggle to meet computational demands efficiently because of growing machine learning enthusiasm.

This work aims to reconfigure shared computing facilities to allow for more efficient machine learning inference from their numerous users. In Section~\ref{sec:fnal_example}, we use the Fermilab shared facilities to show how an NVIDIA Triton Inference Server can be deployed and used to optimize machine learning inference when scaled to multiple users running parallel computing jobs. Section~\ref{sec:benchmarking} then shows the computational gain and the effect of optimizing such a configuration at Fermilab. All results are specific to the Fermilab facility, but the tests and trends are reproducible by all similar multi-user facilities and are anticipated to show similar results.

\section{Background}

In this section, we define shared computing facilities and distinguish the different machine learning processors that are typically made available to users. We then briefly discuss the NVIDIA Triton Inference Server and how it interacts with the different processors.

\subsection{Shared computing facilities}

Computing facilities are widely used around the world to share computing resources among users~\cite{cluster_comp}. As computational tasks become more complex and computationally expensive, shared facilities hold great value by allowing users to access powerful machines that are expensive to own individually. A few companies offer services that give the public access to their computing clusters for a fee, such as Microsoft Azure, Amazon Web Services, Google Cloud Platform, and IBM Cloud. Other companies, universities, research collaborations, and federal laboratories maintain private computing facilities to enable their researchers and employees to perform cutting-edge computations with a scope far outstripping the resources that can be dedicated to typical individuals.

Within HEP, researchers need the capability to process data in the terabyte (TB) to petabyte (PB) range, which may represent the sum of collected information for billions of particle physics collisions or years of continuous data collection. Subsets and variations of the data analysis processing may be repeated thousands of times each year. For the LHC experiments~\cite{atlas,cms}, data processing is typically facilitated by large CPU-centric computing clusters like the LHC Physics Center (LPC)~\cite{lpc}. 

\subsection{Common machine learning processors}\label{sec:ml_processors}

Revolutionary advancements in the past decade have enabled machine learning to become a ubiquitous feature in modern research and commercial environments.
As the field continues to develop, many of the resulting algorithms take increasingly larger proportions of the available computing power and runtime.
GNNs, notable for their ability to process irregularly structured graph-like data, are an example of an ML model that can be rather complex and consequently poses a computational burden when processing large sets of data.
GNNs also represent a transformative paradigm shift for HEP, which naturally deals with events containing diverse and irregularly shaped inputs, often without an intrinsic ordering.
Until their advent, HEP data needed to be heavily pre-processed for ML models having regular input shapes, with significant feature engineering involved, to attain high performance; GNNs have enabled similar or better performance with fewer input features, which is very desirable for HEP data.
It is imperative for facilities like those employed in HEP to evolve and adapt to accommodate multiple users running complex machine learning algorithms, in order to avoid decreased computational efficiency and increased costs to researchers both in terms of money and time.

The two most common processing hardware classes seen at shared computing facilities, the CPU and GPU, have different trade-offs for running machine learning algorithms.
CPUs can be faster in data transfer and storage, with better branch prediction and shorter pipelines, all of which are suited to general-purpose workflows. However, they are limited in parallelism and therefore computational throughput. 
Commercial GPUs, being designed for highly parallel paradigms like single instruction multiple data (SIMD) workloads~\cite{MARINESCU202341}, are particularly well-suited to accelerating ML training and inference~\cite{cpu_vs_gpu}. By trading the more complex branch-prediction hardware and low pipeline latencies of CPUs for more vectorized compute capability, these devices gain considerable advantage in total FLOPS and compute/watt. 
GPUs are more expensive than CPUs (an individual NVIDIA H100 costs around \$35,000, whereas a 32-core AMD EPYC 7543 is approximately \$2,350 in 2023), but have $\mathcal{O}(10)$ better performance per watt, which closes the cost gap.
In combination with lower general-purpose utility and need for specialized programming paradigms and code, GPUs are less frequently employed in HEP computing centers. 
Multi-user computing facilities are obliged to allocate such expensive resources efficiently for the increasing fraction of researchers using ML techniques.

A concept frequently considered in HEP is the time-to-insight, which is the amount of time it takes for a new idea to be proposed, implemented, validated, and analyzed on TB to PB data quantities. 
Being able to provide a short, large burst of resources to an analysis has significant benefits to users. 
However, while minimizing analysis latency is paramount, it must be balanced with achieving high computational efficiency in shared facilities. The NVIDIA Triton Inference Server~\cite{triton} supports both of these goals when paired with GPUs to augment multi-user computing facilities.

\subsection{NVIDIA Triton Inference Server}

One way to minimize cost while providing high burst capability is to provide GPUs as centralized resources for offloading ML computations, while general-purpose calculations are distributed across CPU-only servers. 
The GPUs are then accessed on-demand, with usage requests satisfied on the order of seconds, rather than minutes or hours, as is typical when requesting dedicated GPUs at HEP computing clusters.
This paradigm, known as Inference-as-a-Service (IaaS), can be accomplished using the NVIDIA Triton Inference Server~\cite{triton}, which is open-source software that allows users to send inference requests from any framework to any CPU- or GPU-based platform.
With this tool, shared computing facility users can run all of their code on CPUs except for the ML inference, which will take place on a GPU. 
A Triton server can simultaneously handle ML inference requests from multiple users, for multiple models, using multiple ML frameworks such as PyTorch and TensorFlow~\cite{pytorch,tensorflow}.

With the Triton server set up on a cluster of GPUs, multiple models can be accessed in a device-agnostic way. All server instances connect to an object store where ML models are uploaded, and any server can dynamically load any model that a client requests. 
Additionally, dynamic batching can concatenate inference requests with sub-optimal batch sizes, perform the inference with near-peak efficiency by filling the GPU registers, then split and return the results to separate clients~\cite{INOUE2021102183}.
An individual client is not constrained by how many models can fit into device memory locally, and so may address dozens of models in fast succession, taking advantage of a one-to-many client-to-server connection via one unified interface~\cite{triton_docs}.

\section{Fermilab Triton server application} \label{sec:fnal_example}

In this section, we discuss examples of shared computing facilities at Fermilab and how the NVIDIA Triton Inference Server is deployed.

\subsection{Computing facility statistics}

Fermilab is a national laboratory in the United States which specializes in particle physics research. It is the host laboratory of the US CMS Collaboration, which studies the fundamental particles of the universe using the CMS detector located at CERN in Geneva, Switzerland~\cite{cms}. As such, Fermilab has several computing clusters accessible to US CMS researchers for all their computing needs.

Two shared computing facilities at Fermilab used in this work are the LHC Physics Center (LPC) and the Elastic Analysis Facility (EAF). The LPC is reserved for US CMS-affiliated researchers and has 240 cores available for interactive use (via 60 virtual machines) and another 4500 cores for batch submission. Each LPC batch node has a 10 Gb/s ethernet connection. The LPC currently has hundreds of users and is predominantly used for data analysis. The EAF is also designed for physics analysis, but is accessible to any Fermilab affiliate, intended to provide industry-standard data science frameworks and toolkits for low-latency analyses. It is built on the OKD~\cite{okd} framework (the community-supported distribution of Red Hat OpenShift~\cite{openshift}), which provides scalable, reliable, multi-tenant Kubernetes~\cite{kubernetes}. The EAF consists of 12 machines with 286 CPU cores and 1643 GiB of memory, along with 8 NVIDIA A100 80 GB GPUs. It can also submit large workloads to the LPC batch system.

\subsection{Typical user workflow}

\begin{figure}
    \centering
    \includegraphics[width=0.65\textwidth]{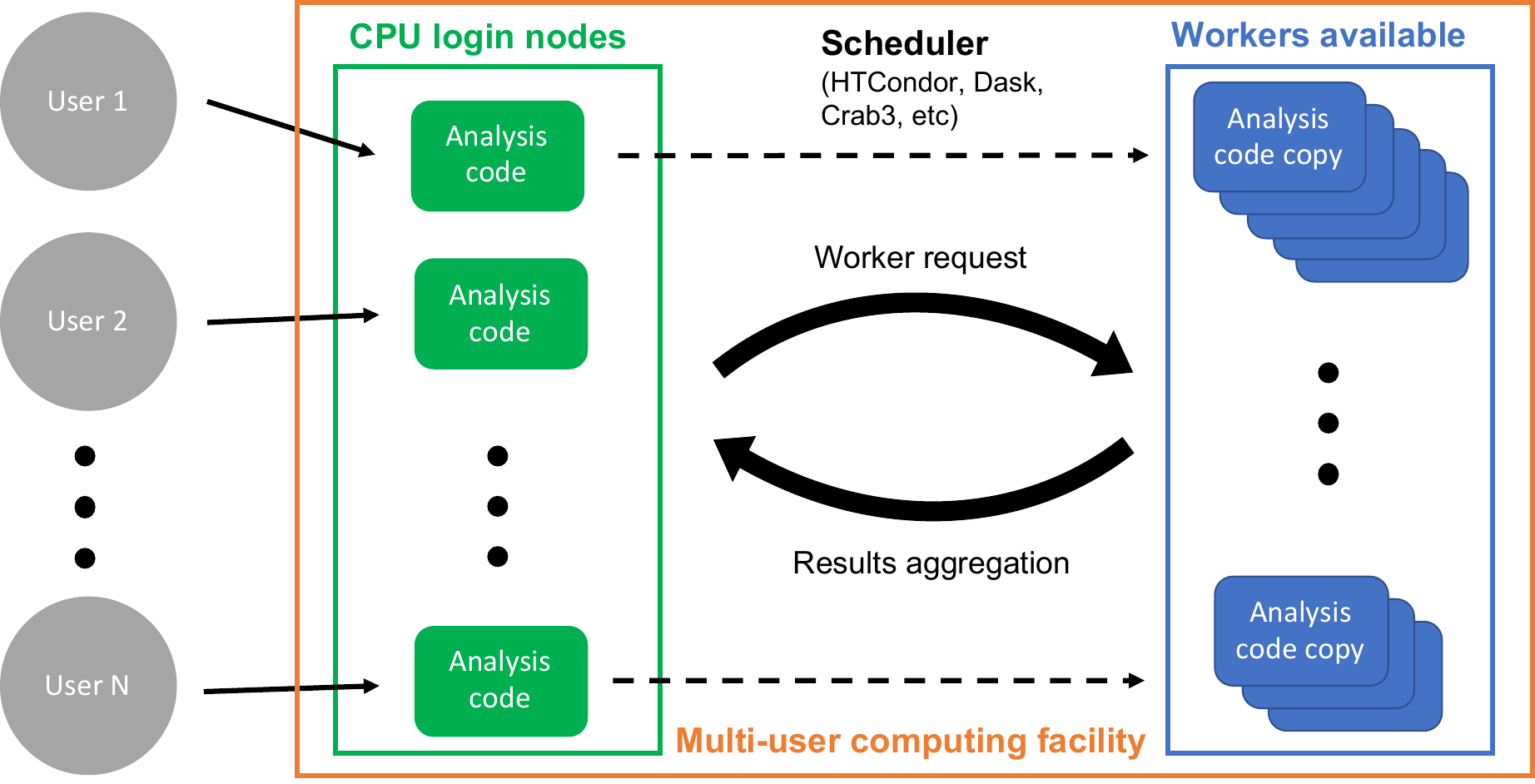}
    \caption{A schematic of a typical user workflow at a shared multi-user computing facility. This example is based on the LPC facility.}
    \label{fig:lpc_workflow}
\end{figure}

Users at the LPC and EAF typically use these computing facilities for data analysis. Upon connecting to one of these facilities, the user will be assigned to a node with access to communal software and storage areas. The collaborator then processes things in two ways: either locally on the login CPU node, or by distributing units of work to multiple CPUs/GPUs through a job scheduler. Figure~\ref{fig:lpc_workflow} shows this typical user workflow as a schematic.

Physics analyses generally entail running the same code over billions of physics events. The analysis code is structured for immense parallel processing over the many data files storing all of these events. Therefore, users generally package a copy of their code to send to each CPU/GPU along with different subsets of the data to analyze so that total processing time is minimized. There are a number of tools that are used to scale the code out within large computing clusters, such as HTCondor~\cite{htcondor}, CRAB3~\cite{crab}, and Dask~\cite{dask}.

Machine learning algorithms are becoming more commonly used by physics data analysts for a variety of tasks, such as event reconstruction and object classification~\cite{ml_in_hep}. 
When running analysis code on CPUs, machine learning inference generally takes up a significant amount of the full processing time, depending on the model. 
Utilizing GPUs to process the entire analysis would speed up the inference time, but is not efficient as significant portions of analysis code are not adapted for GPU usage.
Optimizing this efficiency is imperative when the demand for GPUs exceeds what is available, as is the case in many computing facilities.

\subsection{Triton server implementation}\label{subsec:iaas_impl}

Instead of running uniquely on a CPU or GPU, the Triton server allows these two processors to work together. GPU resources are allocated to the server, which uniquely identifies each available model and dynamically loads needed models so that CPU clients can communicate inference requests. The researchers then send copies of their code to CPUs that compute everything locally except for the ML inference, which is processed by the GPUs on the Triton server. This implementation allows for fast GPU ML inference shared among multiple users.

\begin{figure}
    \centering
    \includegraphics[width=0.8\textwidth]{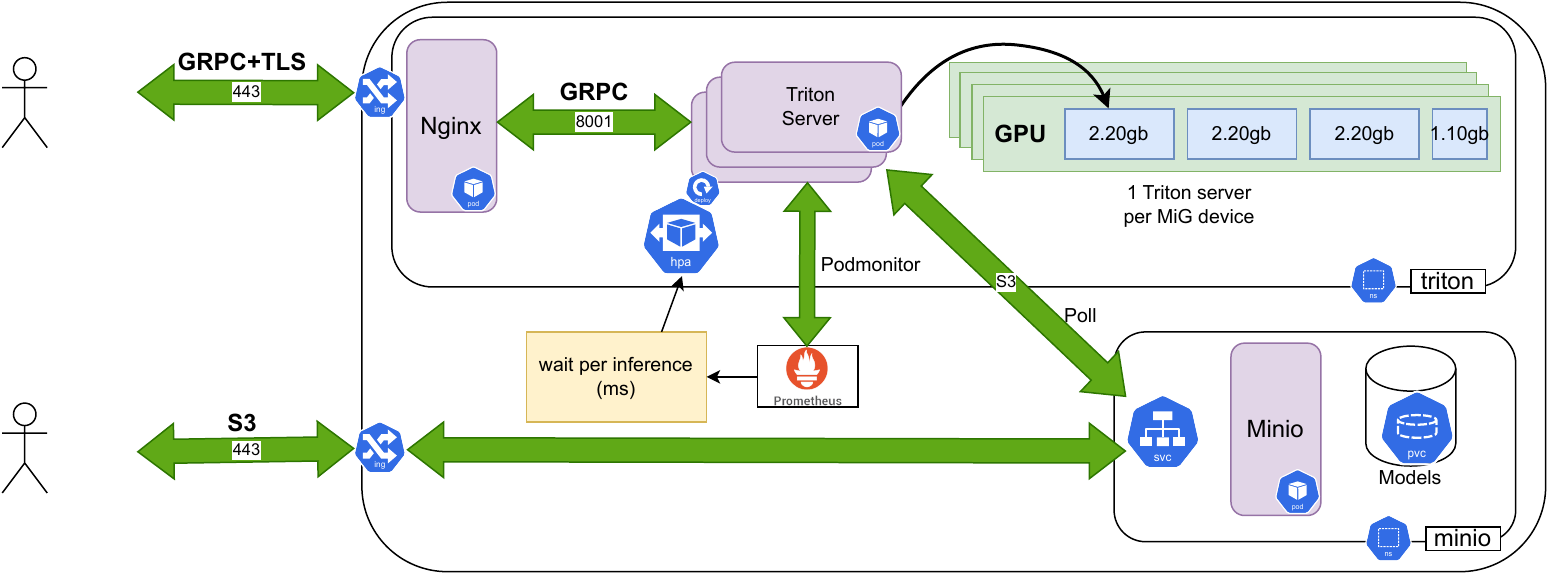}
    \caption{The Triton server implementation at the EAF showing the path of an ML inference request as it is created by the user and processed by the servers.}
    \label{fig:impl_diagram}
\end{figure}

A diagram of the current implementation is shown in Fig.~\ref{fig:impl_diagram}.
It includes two inference machines, each with 4 NVIDIA A100 80 GB GPUs and 2 AMD Epyc 7543 32-core CPUs.
The Ampere architecture's Multi-Instance GPU (MIG) capability is utilized to partition the GPU resources into multiple virtualized resources, and Triton Inference Server instances are deployed on MIG slices with 20 GB of RAM and 14 Streaming Multiprocessor (SM) cores. In Section~\ref{subsec:multimodel}, we also deploy MIG slices with 40 GB of RAM and 28 SM cores.

The A100 architecture has 1935 GB/s of bandwidth to the High Bandwidth Memory attached to the die, and 6912 CUDA cores, providing up to 19.5 TFLOPS of compute on FP32 data and 9.7 TFLOPS for double precision FP64. Each MIG slice has dedicated L2 caches, DRAM bandwidth, and memory controller allocations, helping ensure consistent performance regardless of the usage of neighboring MIG slices.
Each Triton server periodically polls a MinIO~\cite{minio} object store where all the models are stored.

Inference requests originate from worker nodes on the LPC batch system. 
Users send requests via TLS-wrapped gRPC~\cite{grpc} to a haproxy~\cite{haproxy} service built into OKD (not pictured), which are then immediately passed through to an nginx~\cite{nginx} service.  
The nginx service unwraps the gRPC request and sends it to a Triton Inference Server, using Kubernetes load-balancing.
The Triton Inference service is configured to automatically scale up and down the number of server instances based on the average queue time for an inference request (called ``auto-scaling''). 
Each inference machine is connected via 100 Gbps ethernet; however, the nginx and haproxy servers are only connected to the fabric of the LPC batch system at 100 Gbps. This connection could be a bottleneck when numerous LPC batch workers are making inference requests.

The Prometheus open-source monitoring system~\cite{prometheus} built into OKD is used to collect inference metrics every 15 seconds from the Triton application via Kubernetes podmonitor objects, as well as machine characteristics such as core/memory utilization.
The metrics are written to a Grafana Mimir~\cite{grafana} server for long-term storage, accessed via the REST API, and displayed via Grafana monitoring. The metrics collected by Prometheus are used to analyze the performance of the system in Section~\ref{sec:benchmarking}.

\subsubsection{Parameter optimization}\label{sec:param_opt}

\begin{wrapfigure}{r}{0.5\textwidth}
    \vspace{-1cm}
    \centering 
    \includegraphics[width=0.49\textwidth]{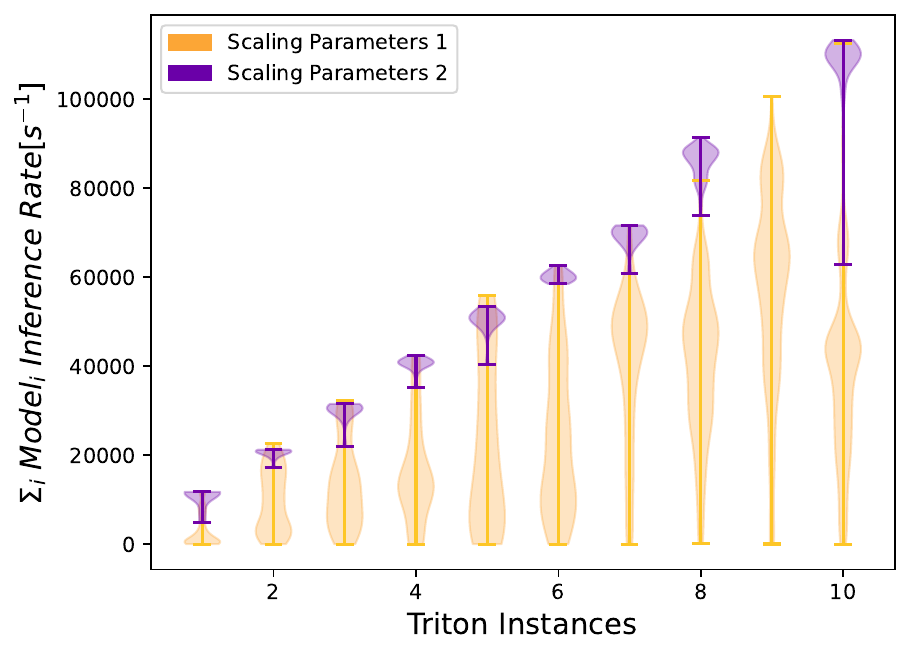}
    \caption{The violin plots show the net inference rate (inferences/s) as a function of the active Triton instances for two different sets of scaling parameters. Each violin shows the minimum, maximum, and (through the width of the shaded band) the frequency of time samples (120 s long). For SP2, the server scaling skips from 8 to 10 instances as additional GPU resources became free.}
    \label{fig:parameter_opt}
    \vspace{-.2cm}
\end{wrapfigure}

Multiple free parameters must be chosen when deploying the Triton server, which affects how quickly and efficiently models can be processed given the resources allocated. The parameters associated with the EAF Triton implementation mentioned above are all based on a standard GNN model used frequently for HEP applications, ParticleNet~\cite{Qu:2019gqs}. The ParticleNet GNN applies dynamic graph convolutional neural networks and edge convolution techniques to variable-dimensioned, unordered ``point cloud'' data. This model (exact model parameters given in App.~\ref{app:particlenet}) will be used as the demonstration model in Section~\ref{sec:benchmarking} and is a fair representation of the ML models being used in HEP today.

The size of the MIG slice (20 GB) for a server instance was chosen based on the RAM required to execute inference requests on the ParticleNet model. Section~\ref{subsec:multimodel} will discuss how performance changes as this parameter varies.

The queue time per inference request is sampled every 15 seconds. If the average queue time exceeds 400 ms for four consecutive samples and it has been at least 3 minutes since the last scale up, an additional server is deployed. Conversely, if the average queue time is less than 400 ms for 40 consecutive samples and it has been at least 1 minute since the last scale down, a server is shut down. These settings are collectively referred to as Scaling Parameters 2 (SP2). The pre-optimized scaling parameters (Scaling Parameters 1, SP1) used a 100 ms threshold on the average over all models and different windows for scaling up and down. See App.~\ref{app:triton} for more scaling parameter information.

Figure~\ref{fig:parameter_opt} depicts the throughput of the ParticleNet model at the EAF for SP1 and SP2. The naive expectation is linear scaling of the maxima as a function of instances. With pre-optimized parameters SP1, the servers are under-utilized, with unused inference capacity the majority of time. Post-optimization gives the performance seen by SP2, demonstrating larger throughput and more consistent scaling with respect to Triton instances, which better maximizes the per-GPU throughput with the ParticleNet model. This indicates the importance of proper parameter selection.

\section{Benchmarking tests}
\label{sec:benchmarking}

In order to understand the benefits of setting up an NVIDIA Triton Inference Server at a shared computing facility, a few metrics are computed and analyzed. 
The timing and computational efficiency for the setup described in Section~\ref{sec:fnal_example} are assessed. 
While all of the results shown in the subsections below are specific to the LPC and EAF Triton server setup at Fermilab, these tests can also be used as benchmarks for other Triton server deployments (code is publicly available at \url{https://github.com/cgsavard/triton_multiuser_benchmarks}).

\subsection{Timing comparison}\label{subsec:timing}

\begin{wrapfigure}{r}{0.5\textwidth}
    \vspace{-2cm}
    \centering 
    \includegraphics[width=0.49\textwidth]{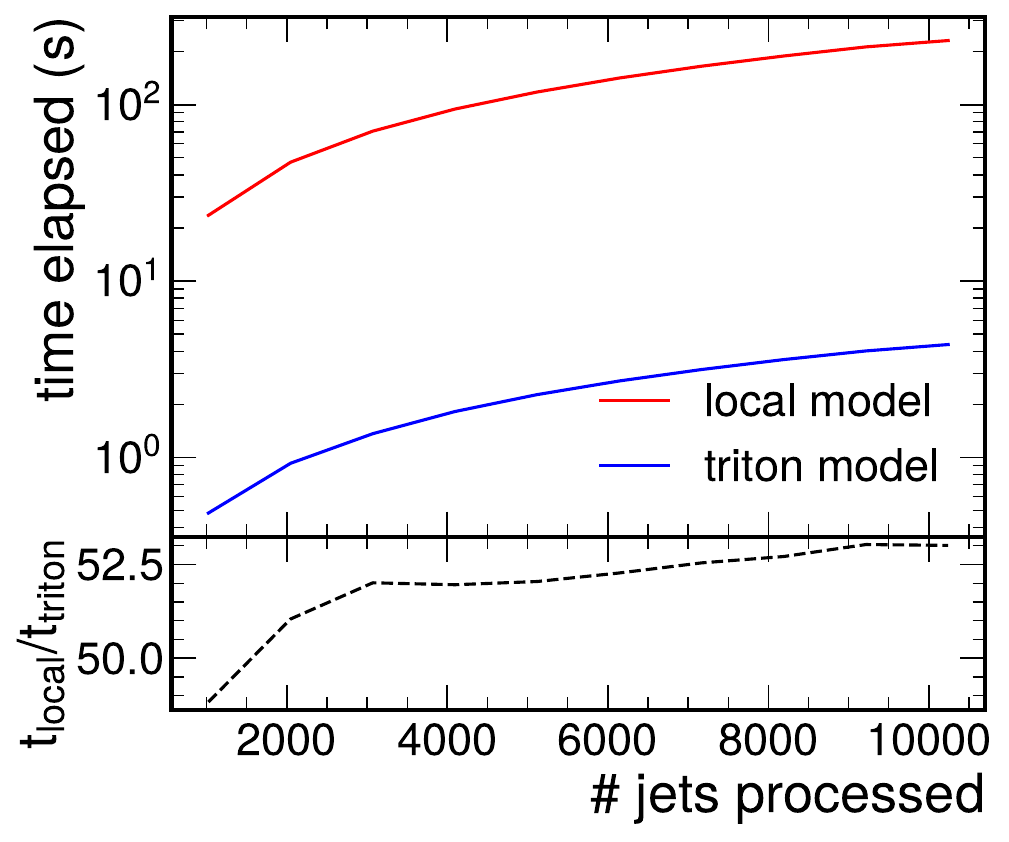}
    \caption{Comparison of the time it takes to process batches of data using a local CPU model vs. a Triton model instance on a GPU for ParticleNet.}
    \label{fig:timing}
    \vspace{-.1cm}
\end{wrapfigure}

At the LPC, users typically run their ML models on the CPU nodes readily available to everyone. 
As discussed previously in Section~\ref{sec:ml_processors}, CPUs are not as efficient for machine learning inference as GPUs.
The Triton server setup, which allows users to execute their inference on a GPU, therefore greatly reduces the overall computing time. 
For this test, we compared the processing time for inference on a local CPU instance of the ParticleNet model to a Triton instance of the model hosted on GPUs.

Figure~\ref{fig:timing} shows a significant speed-up of $\mathcal{O}(50)$ when processing 10,000 inputs (called ``jets'' for the ParticleNet model), motivating the use of the Triton server.
The time elapsed starts when the full dataset is passed to the model and ends when all of the inference results are available, including data batching and pre-processing into the proper format for the selected model. Each data point on the plot represents the time elapsed (cumulative) after processing the indicated number of jets, with the batch size set to 1024.
To minimize noise, which causes small timing fluctuations, the time elapsed is averaged over 10 trials for the local model and 100 trials for the Triton model.
The fluctuations for the Triton model are larger than for the local model because of the network connection between the LPC CPUs and the EAF GPUs, which acts as an additional source of noise.

\begin{figure}[t]
    \centering
    \begin{minipage}[b]{0.48\textwidth}
        \includegraphics[width=\textwidth]{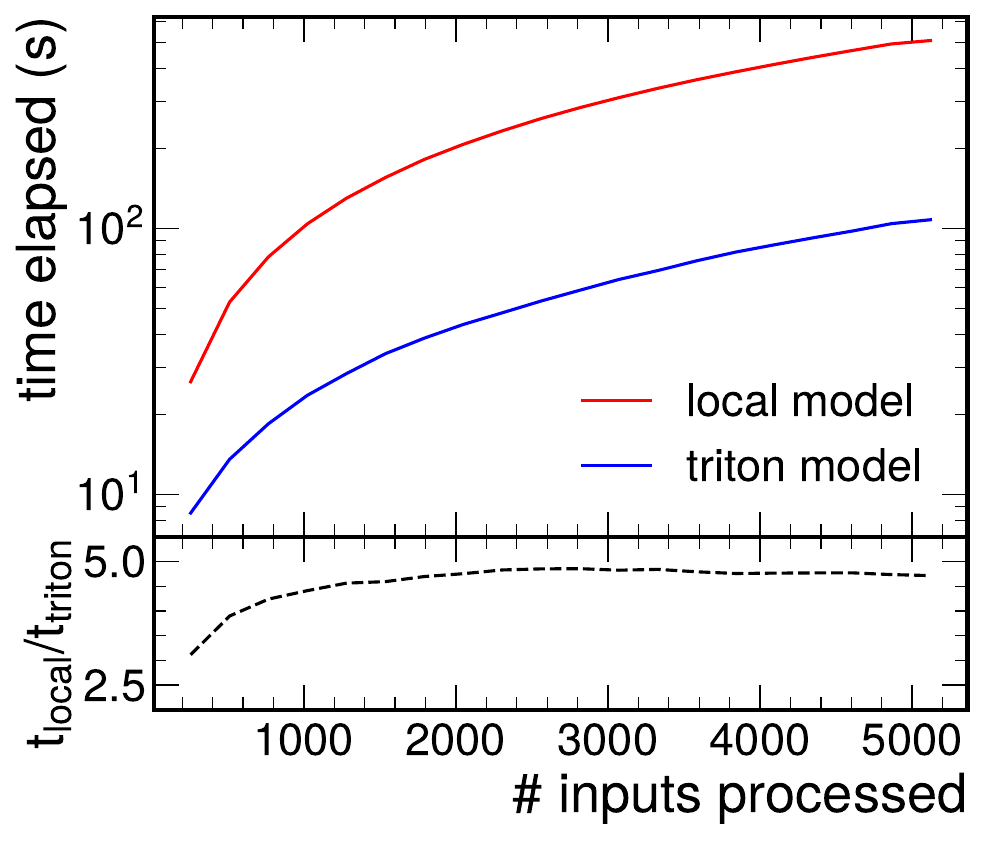}
    \caption{Comparison of the time it takes to process batches of data using a local CPU model vs. a Triton model instance on a GPU for ResNet50. 5000 inputs were processed in batches of 256. Results are averaged over 5 trials.}
    \label{fig:timing_resnet}
  \end{minipage}
  \hfill
  \begin{minipage}[b]{0.5\textwidth}
        \includegraphics[width=\textwidth]{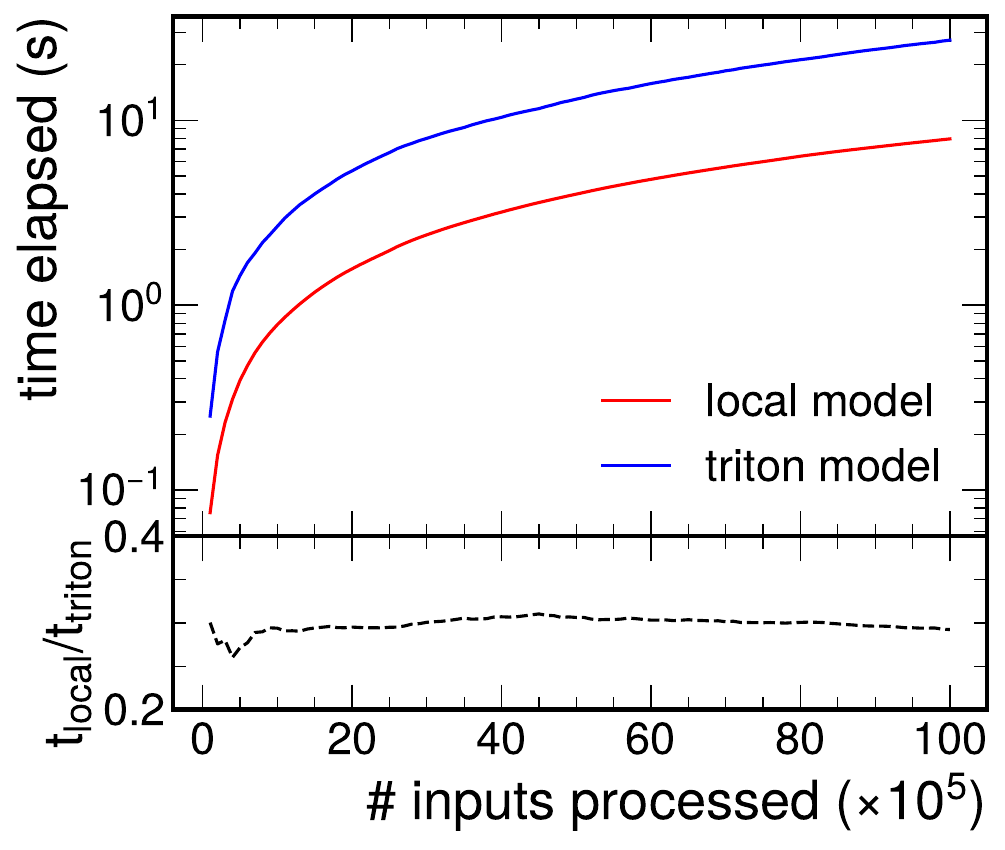}
    \caption{Comparison of the time it takes to process batches of data using a local CPU model vs. a Triton model instance on a GPU for the BDT. 10M inputs were processed in batches of 10000. Results are averaged over 5 trials.}
    \label{fig:timing_bdt}
  \end{minipage}
\end{figure}

It is important to note that different machine learning models will achieve different speed-ups, or even slow-downs, when using a Triton server for GPU inference. In Fig.~\ref{fig:timing_resnet}, we can see a speed-up of $\mathcal{O}(6)$ for a ResNet50 model \cite{he2015deep,canziani2017analysis} when using the same Triton set up described in Section~\ref{sec:fnal_example}. ResNet50 has approximately 12 times more parameters and 7 times more FLOPS than ParticleNet \cite{Qu:2019gqs}, as well as approximately 47 times larger inputs. Thus, the inputs of ResNet50 are a lot larger relative to the size of the neural network in comparison with ParticleNet. This causes the input processing step of the Triton inference to be a much larger fraction of the total inference time, about 10\% compared to $<$1\%. Therefore, the speed-up for GPU inference is smaller, as the input processing is less efficient than inference computation on the GPU.

Figure~\ref{fig:timing_bdt} shows an example of a model that takes more time for inference on the Triton server GPUs than on the local CPUs. This model is a small boosted decision tree (BDT) with 20 input features and 100 trees, trained using XGBoost~\cite{Chen2016}. BDTs are machine learning models that already run very efficiently on CPUs because the inference computation is dominated by simple logical operations. When using the Triton model, there is overhead that stems from data transfer and the packaging/unpackaging of the data. In this case, we see that the overhead from the Triton server masks any speed-up from accelerated GPU computing. Therefore, it is a bad choice to implement the BDT on the server, as it wastes the valuable GPU resources. Users of the server should always test their models to make sure that it is actually beneficial to use the Triton server. More tests comparing ParticleNet, ResNet50, and the boosted decision tree can be found in Appendix~\ref{app:perf_analyzer}.

\subsection{Increasing workers}\label{subsec:inc_workers}

Now, we examine how the Triton server performs as a user runs inference in parallel on multiple workers to speed up the total inference time.
For this test, we spawn varying numbers of workers that make parallel inference requests and see how this affects the inference time with the Triton server auto-scaling (as described in Section~\ref{subsec:iaas_impl}).

\begin{figure}[t]
    \centering
    \begin{minipage}[b]{0.48\textwidth}
        \includegraphics[width=\textwidth]{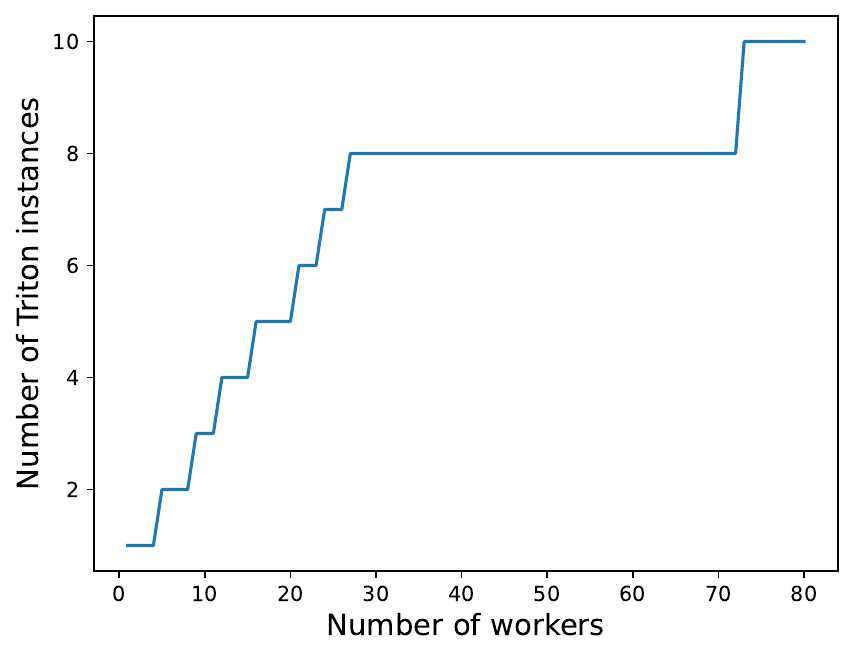}
    \caption{As the number of workers which make parallel requests to the Triton server increases, the number of Triton instances increases to parallelize the request processing.}
    \label{fig:tri_vs_work}
  \end{minipage}
  \hfill
  \begin{minipage}[b]{0.48\textwidth}
        \includegraphics[width=\textwidth]{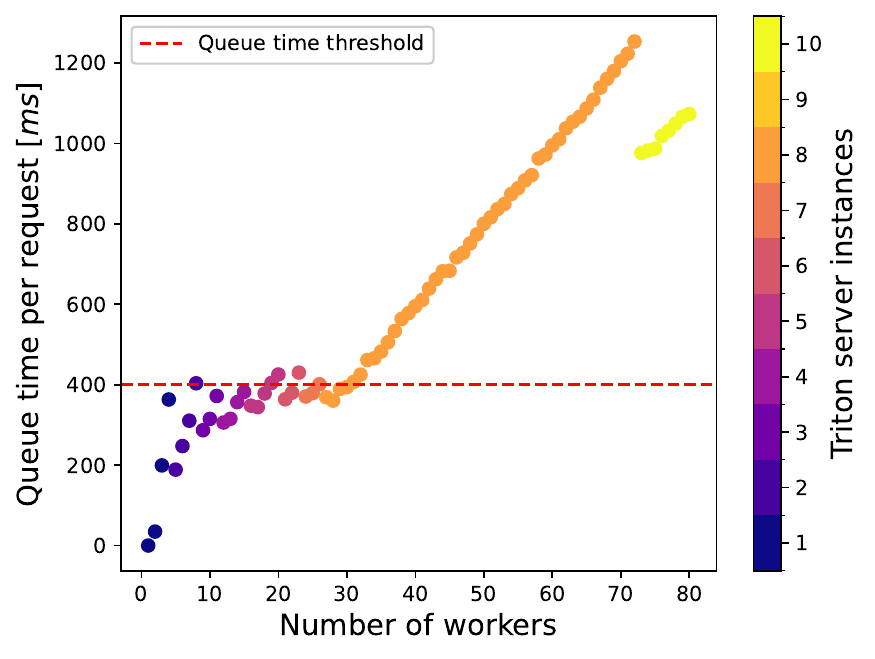}
    \caption{The relationship between the number of workers and the queue time per inference requests, showing the affects of the Triton instance auto-scaling.}
    \label{fig:que_vs_work}
  \end{minipage}
\end{figure}

The Triton instances as a function of the workers can be seen in Fig.~\ref{fig:tri_vs_work}. The increase in instances is steady, determined by the server scale-out rate and queue time threshold, which then remains constant at 8 servers at around 28 workers. As the GPUs on the EAF are a shared resource, no additional MIG slices were available to expand further. Additional MIG slices were freed by other users around the time the benchmark reached 70 workers, and two additional servers were spawned. Fig.~\ref{fig:tri_vs_work} shows how the resources can be reallocated for the Triton server efficiently as more GPUs become available.

Figure~\ref{fig:que_vs_work} shows how the auto-scaling affects the queue time of the requests as a function of the number of workers. A new instance is spawned when the queue time per inference request surpasses the thresholds described in Section~\ref{sec:param_opt}. If the number of instances increases, there are more servers capable of processing requests and therefore the queue time decreases. When the maximum number of instances is reached, the queue will continually increase as more workers send requests and can only decrease when more resources become available to share the load. If the queue time becomes unmanageable because of GPU resource limitations, it may no longer be beneficial to spawn up more workers from the client side.

\begin{figure}[t]
    \centering
    \includegraphics[width=0.49\textwidth]{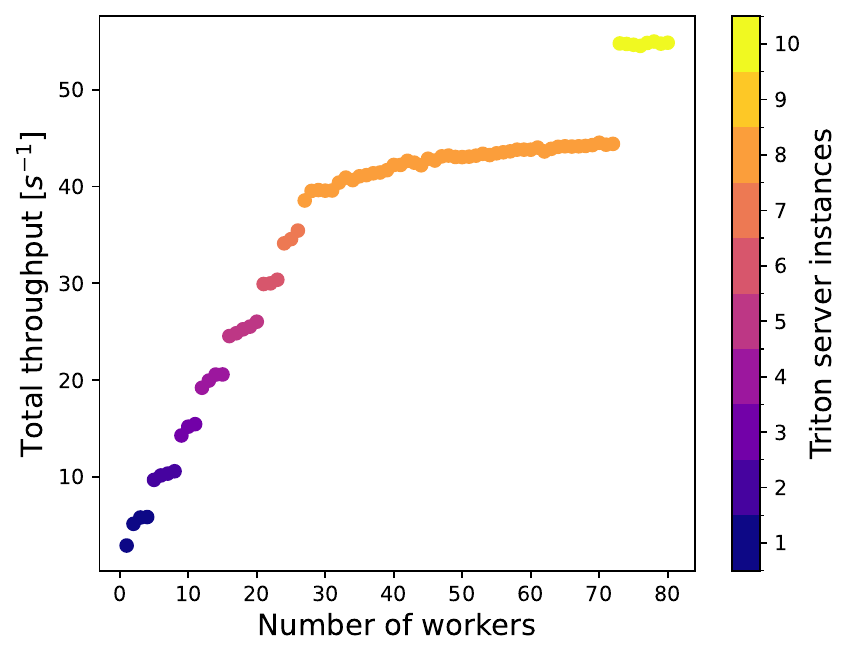}
    \includegraphics[width=0.49\textwidth]{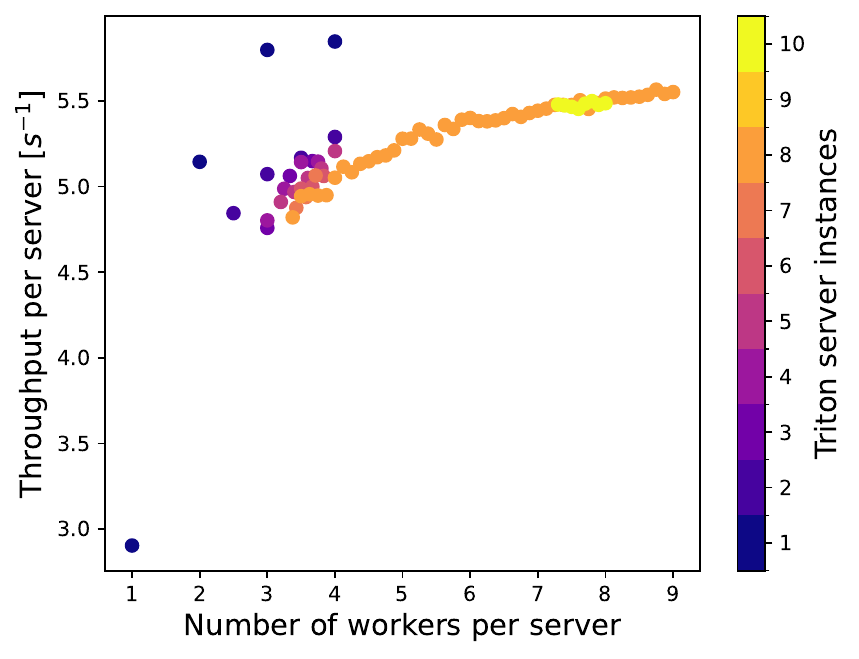}
    \caption{The throughput as a function of the number of workers in the full Triton server system (left) and a single server instance on average (right).}
    \label{fig:reqs_vs_workers}
\end{figure}

The throughput of the Triton server is defined as the rate at which inference requests are processed. As the number of Triton instances increases, more inference requests can be processed in parallel and therefore the throughput increases, as can be seen in Fig.~\ref{fig:reqs_vs_workers}. We may expect the throughput to remain constant so long as the number of servers stays the same, but we actually see a slight increase as more requests fill the queue. The throughput increases as a function of the number of workers because the queuing and processing pipeline becomes more efficient. As the number of instances increases, the processing pipeline stabilizes and the throughput grows more steadily with increases in workers.

\subsection{Multi-model scaling}\label{subsec:multimodel}

In the previous subsection, we looked at the performance of an individual machine learning model using the Triton server for inference. In shared multi-user computing facilities, we expect to have multiple models running inference concurrently. When this occurs, the performance of a single model (``demo model'') can change due to the additional stress put on the Triton server.

\begin{figure}[t]
    \centering
    \includegraphics[width=0.5\textwidth]{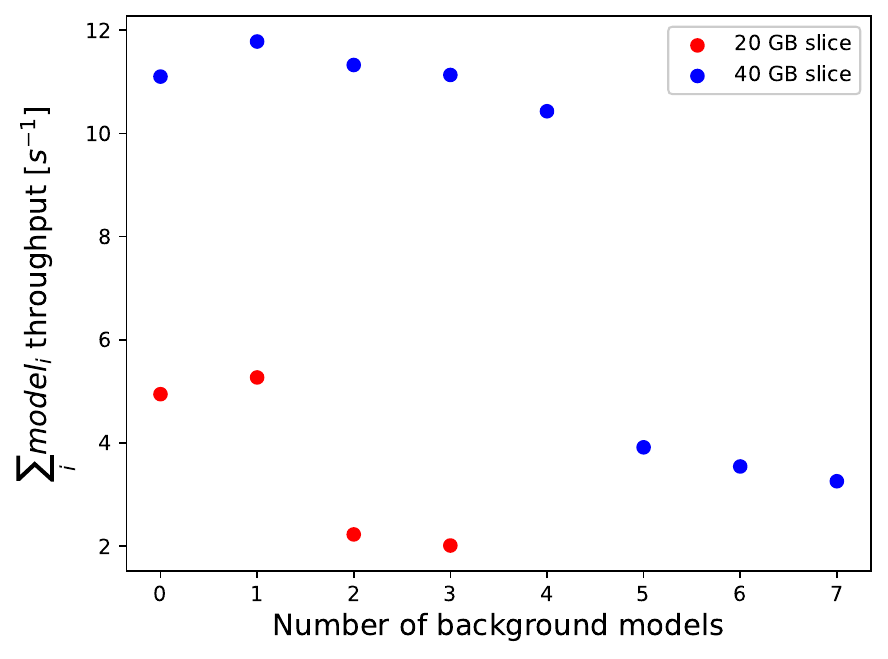}
    \includegraphics[width=0.49\textwidth]{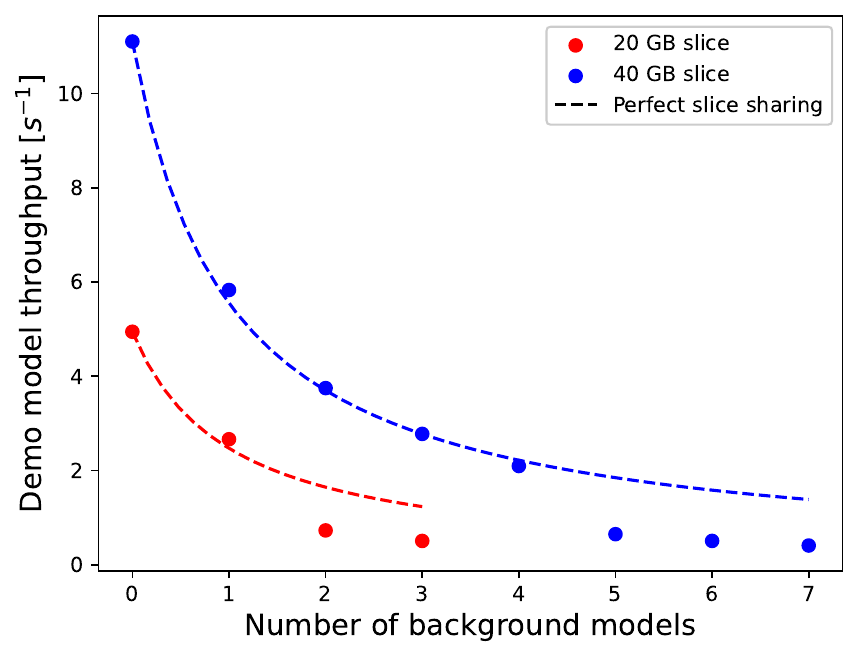}
    \caption{The throughput for all models aggregated (left) and the demo model (right) as a function of the number of additional background models running in parallel with the demo model on a single Triton server. Perfect slice sharing leads to a $1/n$ decrease in throughput with the number of background models $n$. Each model has four workers sending inference requests in parallel.}
    \label{fig:mulmodel_1instance}
\end{figure}

The Triton server loads every model on every server instance running by default. This means that the 20 GB MIG slice hosting an instance is split among the different models and therefore the throughput for a single model decreases. 
In order to test performance when inference occurs for different models at the same time, we created ``background models'': copies of the demo ParticleNet model, but labeled in such a way that the server would treat them distinctly.
Figure~\ref{fig:mulmodel_1instance} shows the relationship between the throughput of all models and throughput of a single model as a function of the number of background models for 20 and 40 GB slices.

The throughput of the individual models scale as $1/n$ when $n$ models are perfectly sharing the GPU slice, as long as there is enough memory for each model to run in parallel. As the number of background models increases, however, the models begin to compete for the instance resources and the throughput decreases faster than $1/n$. This degradation of performance can be due to models loading and unloading on the server or models remaining idle until memory for inference is made available (called ``thrashing''). Figure~\ref{fig:mulmodel_1instance} shows that this thrashing occurs after 2 models on a 20 GB slice and 5 models on a 40 GB, indicating that the demo model requires around 7 to 8 GB minimum in order to run inference efficiently.

\begin{figure}
    \centering
    \includegraphics[width=0.5\textwidth]{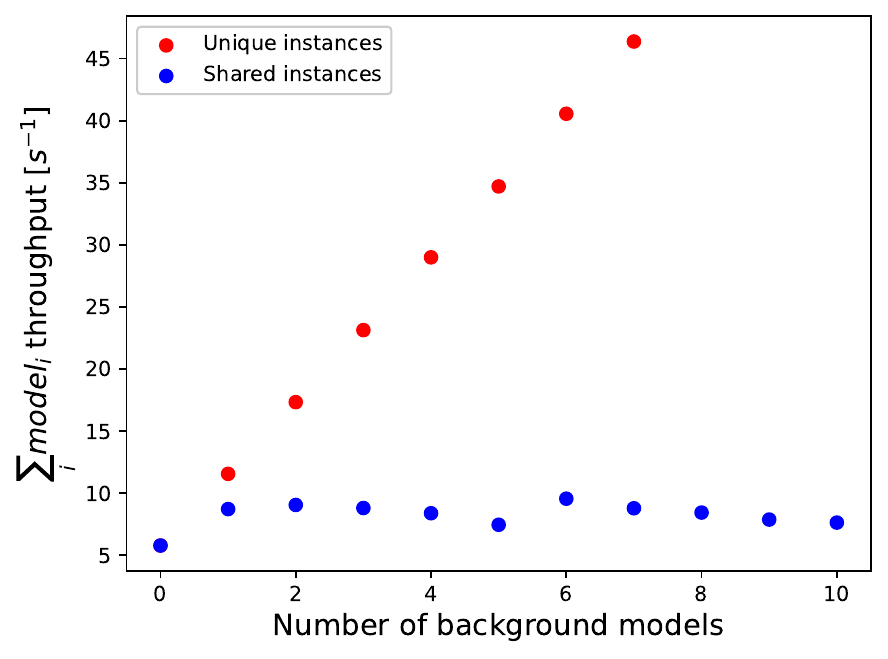}
    \includegraphics[width=0.49\textwidth]{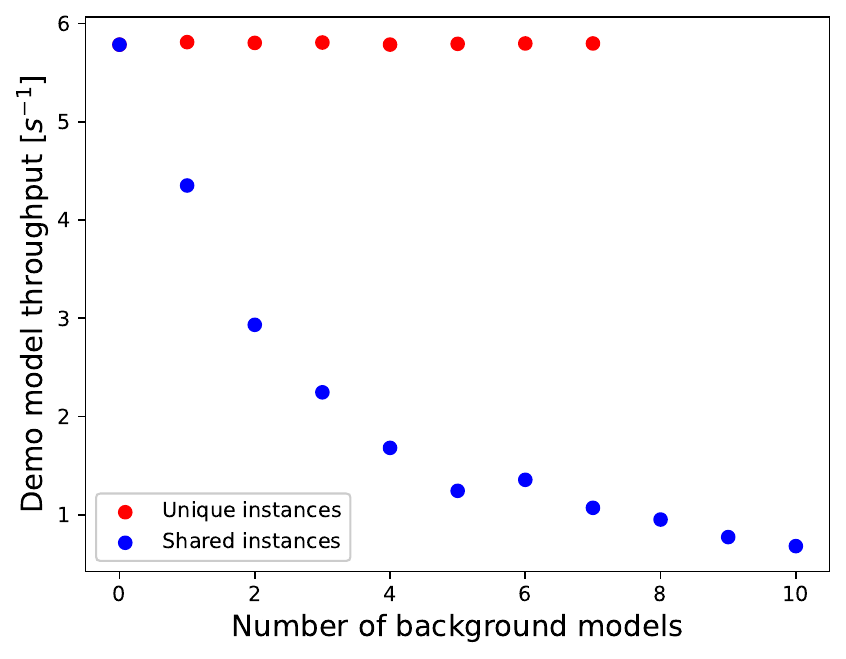}
    \caption{The throughput for all models aggregated (left) and the demo model (right) as a function of the number of background models running in parallel with the demo model with Triton instance auto-scaling. The distributions show differences in performance when each model is hosted on a unique instance versus the models sharing each instance. Each model has four workers sending inference requests in parallel to it.}
    \label{fig:mulmodel_mulinstance}
\end{figure}

Since all models are loaded onto each Triton instance by default, adding more instances does not fix the thrashing that occurs on a single instance. Instead, it is more efficient to make use of the multiple GPU slices available to process each model on a unique instance. Figure~\ref{fig:mulmodel_mulinstance} shows the difference in throughput when all models are sharing each instance versus each instance holding only one model.

The constant throughput of the demo model for uniquely-assigned instances when other models are running in the background shows that processing performance of one model will not affect any other model at a multi-user facility as long as enough GPU resources are available. When GPU resources are constrained, the instances will have to begin splitting among the models carefully to avoid the thrashing seen in Fig.~\ref{fig:mulmodel_1instance}. Such model orchestration is not a feature of the current default Triton deployment and therefore must be implemented manually. NVIDIA's Triton Management Service (TMS), part of their AI Enterprise product, advertises that it "allocates models to individual GPUs/CPUs, and efficiently collocates models by frameworks". Unfortunately, TMS is not currently available to test at the EAF.

\section{Limitations}\label{sec:limitations}

All results mentioned in Section~\ref{sec:benchmarking} are specific to the Triton server implementation at Fermilab described in Section~\ref{sec:fnal_example}. Other facilities may require a different configuration based on the resources available and design of the facility. Similarly, the Triton server parameters optimized in Section~\ref{sec:param_opt} are tuned on a model architecture frequently used in HEP research at the Fermilab facility. These parameters were only optimized on one variant of the model architecture and may need re-tuning as the collection of models used in the multi-user computing facility change or when the Triton server is implemented at a new facility with different network and compute resources.

The benefits of this work, mainly the exploitation of GPUs for quick bursts of resources resulting in high inference throughput and fast turnaround, may be less obvious at a shared computing facility with fewer users or more GPUs. This work also does not explore implementations of inference-as-a-service on different coprocessor architectures such as Tensor Processing Units and FPGAs. These may provide complementary benefits through their performance~\cite{bench_forDL,fpga-a-a-s}, and are an interesting area for additional study and comparison.

These results do not study the potential impact of insufficient GPU resources in detail, leading to over-subscription and untenable latency for the pool of users, nor potential fallbacks in the event that the GPU resources become unavailable for long periods of time.

\section{Conclusion}

In this work, we explore the usage and optimization of NVIDIA Triton Inference Servers at a shared multi-user facility aimed at maximizing throughput when scaling computational resources out to hundreds of users each parallelizing computing jobs. The Fermilab computing facilities have these large-scale computing requirements and are used to demonstrate the performance of model inference-as-a-service under such intense conditions.

The timing comparisons shown in Section~\ref{subsec:timing} motivate using the Triton server to process inference requests on GPUs with a speed-up of ${\sim}$50 compared to CPU-only processing for the ParticleNet model. Sections~\ref{subsec:inc_workers} and~\ref{subsec:multimodel} show how machine learning inference performance on the Triton server changes as parallel requests and active background models increase. Both of these results show that high throughput can be maintained as more stress is placed on the Triton service when the server GPU resources are divided efficiently among the models to maintain a reasonable queue time and minimize competition for resources.

As machine learning becomes more established and ubiquitous in a variety of fields, it is more and more important to ensure that computing centers are capable of handling the increased load from machine learning inference. At shared computing facilities, resources must be allocated to users efficiently, and high throughput is important so that allocated resources can be freed up quickly for use by other users, and the time to insight can be minimized. Triton servers have been shown to efficiently allocate GPU resources for high throughput computing, making this work a leading example of how other multi-user computing facilities can alter their systems to optimize efficiency for new machine learning demands.


\begin{appendices}

\section{Triton server parameters}\label{app:triton}

Several scaling parameters must be set to determine how the Triton server will create new instances, as discussed in Section~\ref{sec:param_opt}. The parameters are carefully tuned to ensure that the instances are scaling out in a stable and efficient manner, as shown in Fig.~\ref{fig:parameter_opt}. These parameters will be described below, along with a brief explanation of how we chose the parameters for the FNAL Triton server implementation.

Each parameter is set uniformly for all models running on the server. These are not configurable on a per-user basis, as any change will affect all users and models using the same Triton server deployment.

\subsection*{Metric collection and analysis parameters}

There are several time-related parameters for the inference server metrics. 
\begin{itemize}
  \item Metric collection interval: the default Triton server settings are used, such that statistics for model inference are collected every 15 s
  \item Analysis time step: sets the interval between analyzed data points. For this analysis, the time step is set to the same value as the collection interval, 15 s.
  \item Data collection window: determines the typical number of metrics used as input for the calculation of rates, deltas, and averages. By selecting an interval of 30 s, 2 consecutive measurements are used to compute an analysis data point. When used in conjunction with a smaller analysis time step, the result is a sliding-window algorithm. This is well suited to averages and queue times.
\end{itemize}
Some metrics, such as the integrated number of requests, must be computed on unique values, and in such a case, the analysis time step and data collection window should be set to the same value to avoid double-counting.
Inference metrics, such as inference request rate and queue time, are calculated and used to determine the performance of the server and whether more instances should be launched or shut down. 

\subsection*{Horizontal Pod Autoscaling Parameters}

The Triton server is configured as a Horizontal Pod Autoscaler (HPA) in Kubernetes~\cite{kubernetes:hpa}. It is configured to scale based on an external metric, referred to as the ``queue time'', which is the maximum of the approximate queue time per inference, averaged per model.
This metric gives a measure of the latency for a single request to be processed in the inference queue. Our implementation chose a threshold of 400 ms, which achieved a smooth scaling of MIG instances while maintaining a reasonable throughput of approximately 5 inference requests per second per instance for the ParticleNet demo model. Note that the throughput of a model is model-dependent and the threshold may need to be be adjusted to achieve reasonable throughput depending on the models being served.

The HPA scaling behavior parameters are summarized in Table~\ref{tab:policyTable}. Given the relatively small amount of MIG instances available (10), \texttt{policies.type} and \texttt{policies.value} were set to "Pods" and "1", respectively, to ensure that we would only start or stop a single server at a time.

\begin{table}[htbp]
  \centering
  \caption{Scaling behavior parameters of Triton HPA}
  \label{tab:policyTable}
  \begin{tabular}{lp{1cm}}
    \textbf{Parameter} & \textbf{Value} \\
    \hline
    \texttt{scaleUp} & \\
    \quad \texttt{stabilizationWindowSeconds} & 60 \\
    \quad \texttt{selectPolicy} & Max \\
    \quad \texttt{policies.periodSeconds} & 180 \\
    \quad \texttt{policies.type} & Pods \\
    \quad \texttt{policies.value} & 1 \\
    \texttt{scaleDown} & \\
    \quad \texttt{stabilizationWindowSeconds} & 600 \\
    \quad \texttt{selectPolicy} & Max \\
    \quad \texttt{policies.periodSeconds} & 60 \\
    \quad \texttt{policies.type} & Pods \\
    \quad \texttt{policies.value} & 1 \\
  \end{tabular}
  
\end{table}

%

%

The stabilization window for scale-up (\texttt{scaleUp.stabilizationWindowSeconds}) was chosen to be one minute, or four measurements (15 second interval) collected by the server. This was found to be a long enough time to determine whether the queue time continuously passes the threshold, but short enough to scale up quickly if the number of inference requests increases suddenly.

\begin{figure}[t]
    \centering
    \begin{minipage}[b]{0.48\textwidth}
        \includegraphics[width=\textwidth]{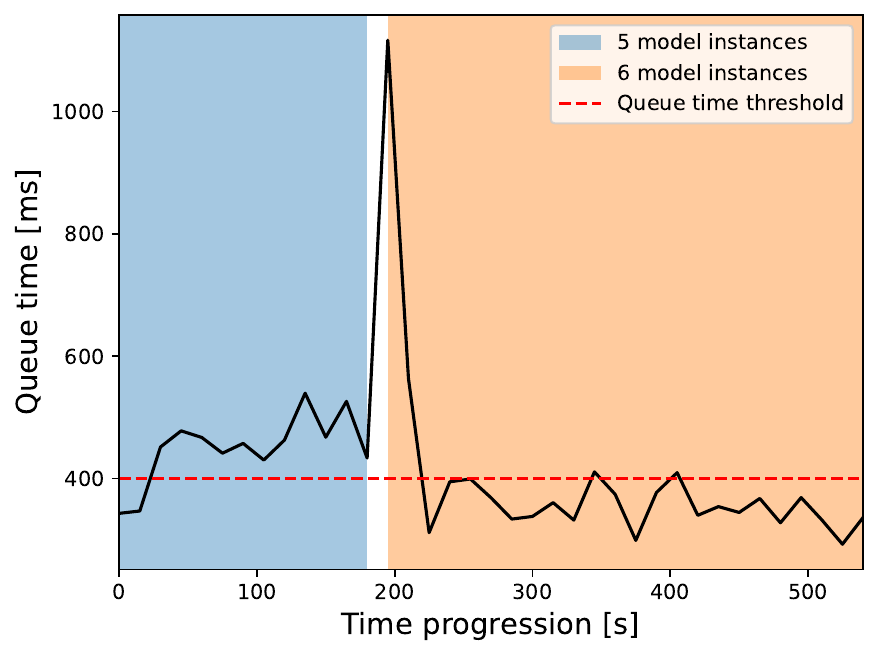}
    \caption{When the Triton server starts up a new model instance, the queue time becomes noisy for a couple of collection intervals until stabilizing again.}
    \label{fig:scaleup_stable}
  \end{minipage}
  \hfill
  \begin{minipage}[b]{0.48\textwidth}
        \includegraphics[width=\textwidth]{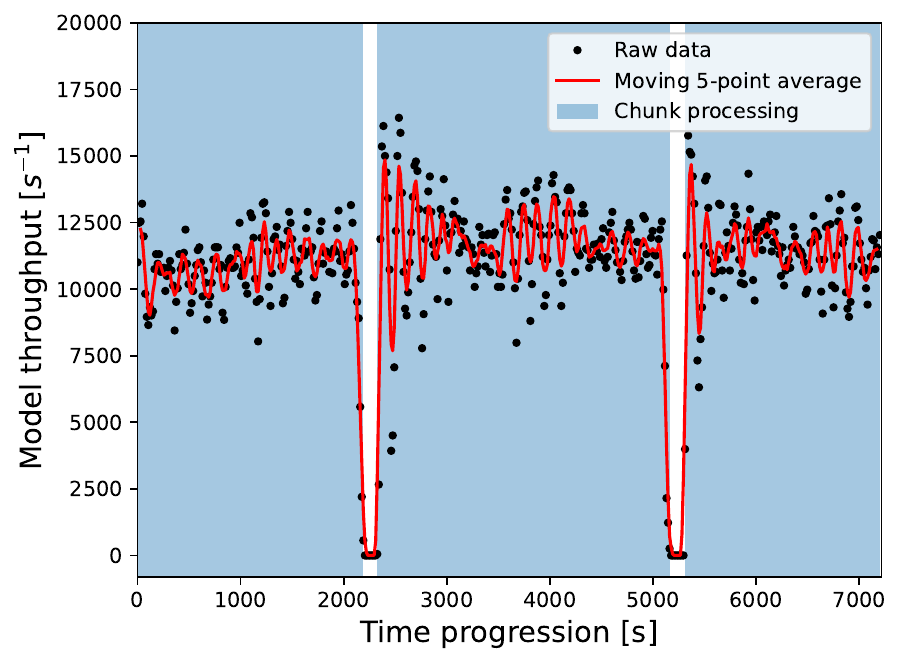}
    \caption{Once a chunk of data is finished processing and while the next chunk is being pre-processed and undergoing non-inference computations, the server throughput drops.}
    \label{fig:scaledown_downtime}
  \end{minipage}
\end{figure}

There is a delay before the queue time responds to a new inference instance being spawned, as seen in Figure~\ref{fig:scaleup_stable}. For this reason, the \texttt{scaleUp.policies.periodSeconds} should be larger than the stabilization window in order to allow the queue time to decrease and stabilize. We chose 180 seconds, allowing the service two minutes for the queue time to stabilize and an additional minute to evaluate if an another instance should be spawned.

To avoid ``flapping'' --- constantly starting and stopping instances as the queue time oscillates around the threshold --- we choose a longer stabilization window for scaling down (\texttt{scaleDown.stabilizationWindowSeconds}): 600 seconds, or 40 consecutive measurements.  This allows for long, uninterrupted processing time.
Since HEP analyses tend to process millions of data events, the data is split into chunks, each generating batches of inference requests directed towards the Triton server. When processing begins for a chunk, a number of synchronous tasks unrelated to inference are performed which causes the number of inference requests to drop to zero until inference for that chunk finally begins, as shown in Fig.~\ref{fig:scaledown_downtime}. We want to ensure that the stabilization window is long enough to avoid scaling down during this downtime, trading off a small inefficiency for a decrease in overall latency.

We set \texttt{scaleDown.policies.periodSeconds} to 60, ensuring that the servers can scale down rapidly once all inference requests have been processed, releasing the allocated resource so they can be used elsewhere. It is beneficial for the \texttt{scaleDown.policies.periodSeconds} to be smaller than the stabilization window as we want stable, long processing times but should be quick to free resources once processing has finished.

\section{ParticleNet demo model parameters}\label{app:particlenet}

The ParticleNet model used in this work to optimize the Triton server parameters and for benchmark testing is an exact replica of the model described in Ref.~\cite{Qu:2019gqs}. There are five input features, two coordinates, three EdgeConv Blocks using $k=16$ nearest neighbors and 
$C=(64, 64, 64)$, $(128, 128, 128)$, and $(256, 256, 256)$ channels, and two fully connected layers with 256 nodes and 0.1 dropout rate to two nodes. A schematic of the exact structure can be found in Fig.~2a of Ref.~\cite{Qu:2019gqs}.

This demo model was developed using the ``Weaver'' package (publicly available at \url{https://github.com/hqucms/weaver}) and was left untrained with randomized weights as the application and performance of the model is unrelated to the performance of the Triton server implementation which we are studying. Similarly, the input data are pseudo-randomized and arranged in the proper format required for inference. The structure of the demo model is based on a ParticleNet model being used in an ongoing physics analysis at the LHC physics center at Fermilab~\cite{lpc}.

The demo model was created using the PyTorch package~\cite{pytorch} and converted using TorchScript to a version that can run on the Triton server. The Triton server reads in the converted file along with configuration files that tell the server how the model inputs and outputs are structured and how to partition the model on the server. In the configuration file, the following selections were made:

\begin{itemize}
    \item \emph{Dynamic batching}: The preferred batch size was set to 1024, determined by testing different batch sizes and balancing the inference speed with memory required to run.
    \item \emph{Inputs and outputs}: There are 3 different inputs for the ParticleNet: the features, coordinates, and mask. Each of these three, along with the output, was assigned \texttt{FP32} datatypes.
    \item \emph{Inference mode}: This is set to \texttt{True}, letting the server know that the model is being used to run inference.
    \item \emph{Instance group}: By default, a single model instance is created for each MIG spawned. This default is kept, as the demo model already requires 7--8 GB for inference and it is easier to test the Triton server implementation with a single model on each MIG. No specific GPU is targeted for the model as the GPUs available to the server are never fixed.
\end{itemize}
All other configurations are left to the default settings.

\section{Nvidia Performance Analyzer}\label{app:perf_analyzer}

NVIDIA's \emph{perf\_analyzer} is part of the suite of Triton tools, designed to generate test inference requests and aggregate metrics. The results may be analyzed to optimize model parameters for inference (see NVIDIA's \emph{model\_analyzer}) and identify model characteristics.

\begin{figure}[ht]
    \centering
    \includegraphics[width=0.31\textwidth]{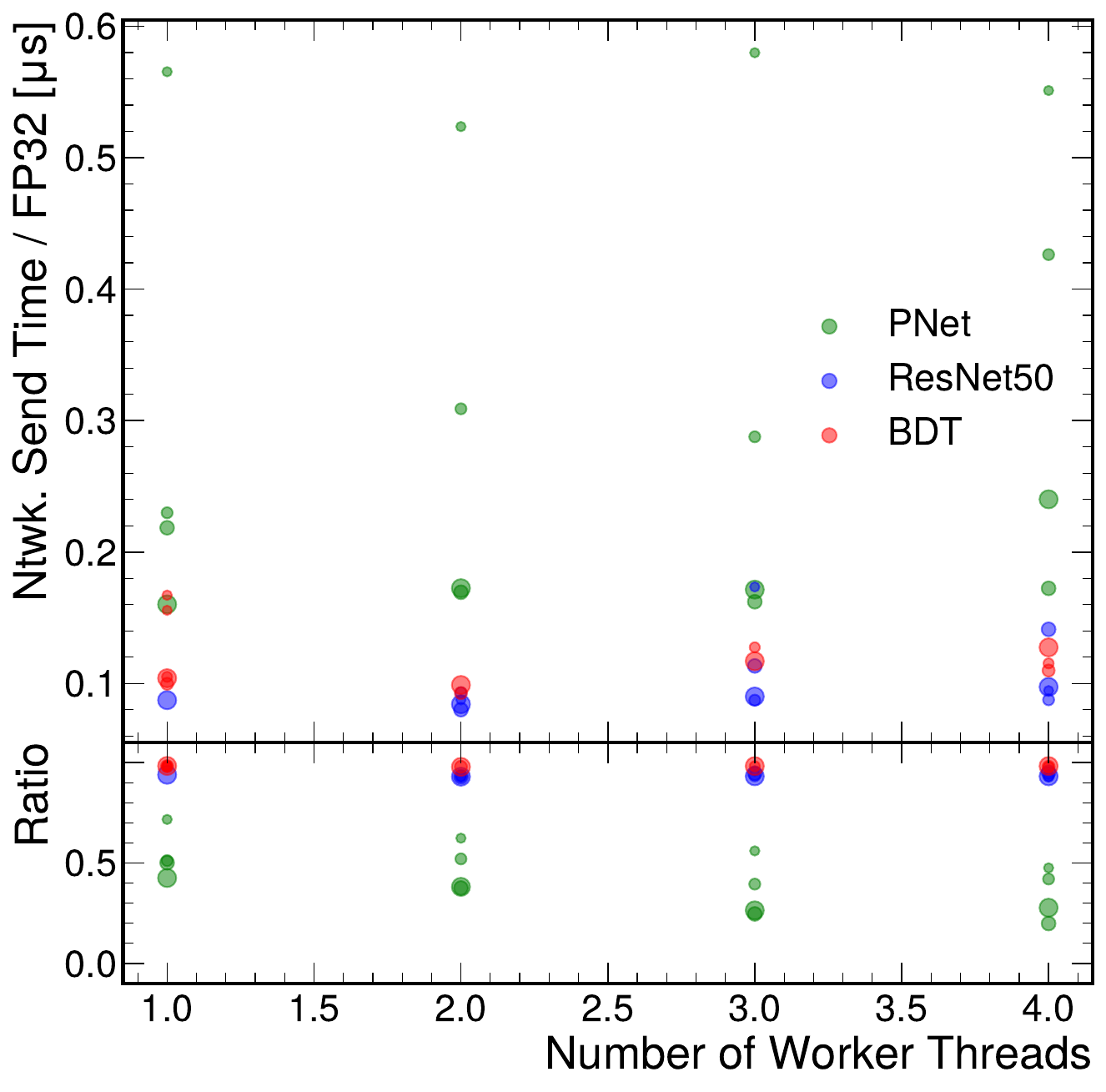}
    \includegraphics[width=0.32\textwidth]{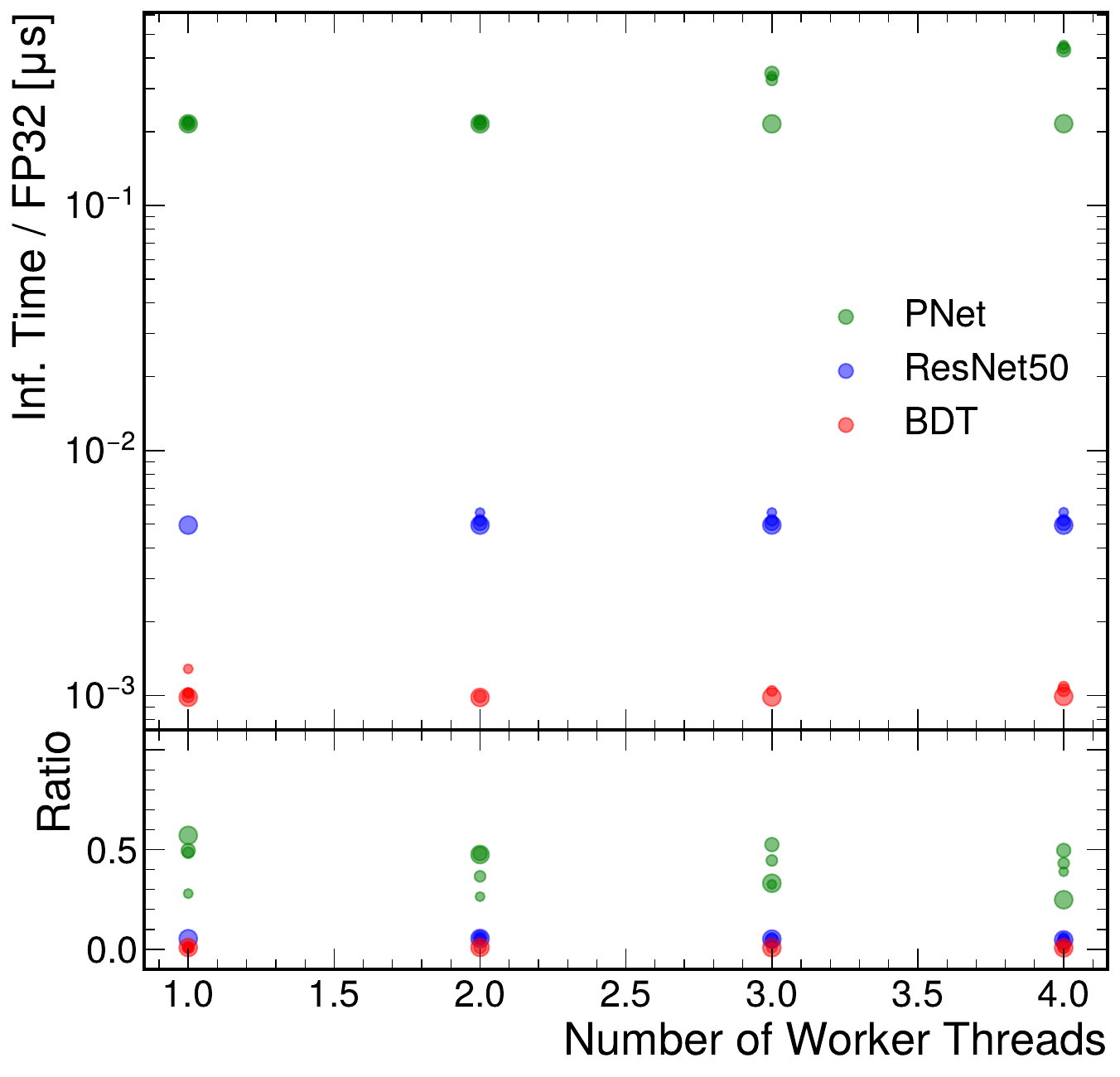}
    \includegraphics[width=0.32\textwidth]{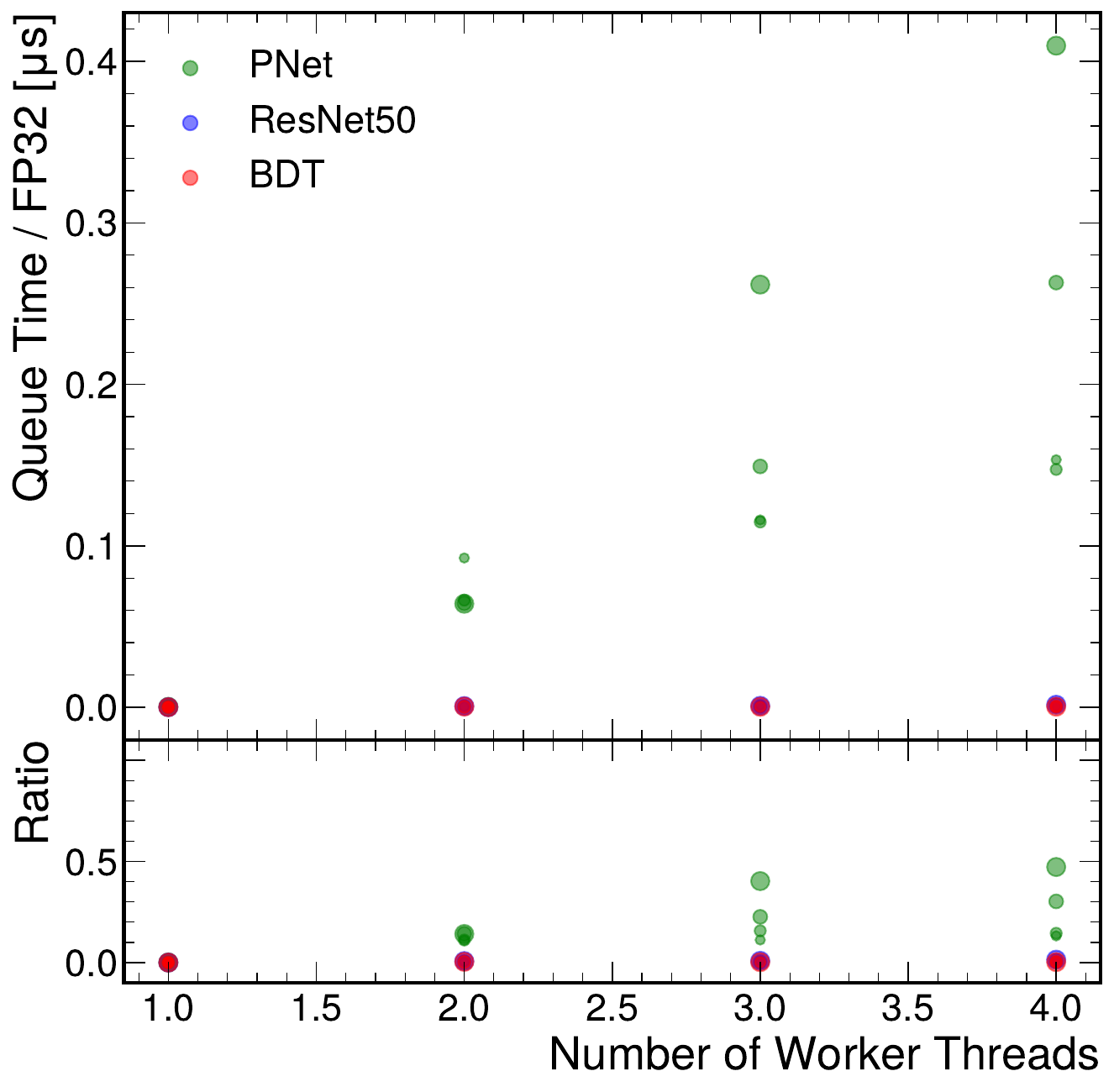}
    \caption{The network latency (left), inference time (middle) and queue time (right) normalized to the number of (FP32) inputs to the model, accounting for batch size. The number of worker threads is how many parallel requests (concurrency) the \emph{perf\_analyzer} on the client machine synchronously sends to the server. The ratios show the fraction of time attributed to each time measure (network latency, inference time, or queue time respectively) to the total request time. The size of the marker corresponds to the batch size, via normalization of the batch size minima and maxima across models.}
    \label{fig:perf_analyzer}
\end{figure}

For a single-instance Triton Server at the EAF, the three models included in this work (ParticleNet, ResNet50, and a boosted decision tree) were benchmarked over a range of concurrencies (the number of simultaneous randomly-generated requests from the \emph{perf\_analyzer}, in synchronous call mode) and batch sizes. The values scanned over can be found in Table~\ref{tab:perf_analyzer}. Fig.~\ref{fig:perf_analyzer} shows the results of these benchmark tests, comparing the normalized latency, inference time, and queue times between the 3 models. The time to send data between the client and the Triton server (network latency) is normalized by the number of 32-bit floating point (FP32) inputs to the model, including the batch size. For example, a BDT with 25 FP32 inputs, batch size of 500, and a total network latency of 14,800 $\mu s$, would have a normalized value of 1.184 $\mu s$. This same normalization is applied to the inference and queue times. Each plot also contains the ratio of the respective non-normalized time to the total inference request time (i.e. the fraction of time which is attributable to the network latency, inference time, or queue time, respectively). For the ResNet50 and BDT models, the total request time is dominated by the network latency. ParticleNet shows non-negligible queue times as the number of worker threads increases, and spends approximately half the time in inference. Models which are more compute-dominated, such as ParticleNet, stand a greater chance of overcoming the network latency overhead of the Inference-as-a-Service model.

\begin{table}[htbp]
  \centering
  \caption{Parameters for \emph{perf\_analyzer} tests.}
  \label{tab:perf_analyzer}
  \setlength{\tabcolsep}{4pt}
  \footnotesize
  \begin{tabular}{lllll}
    \textbf{Model} & \textbf{Batch Sizes} & \textbf{Concurrencies} & \textbf{Input Shape} & \textbf{Output Shape}\\
    \hline
    \texttt{ParticleNet} & 128, 256, 512, 1024 & 1-4 & 8 * 100 & 2\\
    \texttt{ResNet50} & 32, 64, 128, 256 & 1-4  & 3 * 224 * 224 & 1000\\
    \texttt{BDT} & 10K, 40K, 160K, 640K, 2.56M & 1-4  & 20 & 2\\
  \end{tabular}
\end{table}

\end{appendices}
\FloatBarrier



\bmhead*{Acknowledgments}

This work was performed with support of the U.S. CMS Software and Computing Operations Program under the U.S. CMS HL-LHC R\&D Initiative.
This work was partially supported by Fermilab operated by Fermi Research Alliance, LLC under Contract No. DE-AC02-07CH11359 with the Department of Energy, and by the National Science Foundation under grant ACI-1450377 and Cooperative Agreement PHY-1120138. Additional support came from the Department of Energy DE-SC0010005 grant.


\bibliography{sn-bibliography}

\end{document}